\documentclass[a4paper,fleqn,usenatbib]{mnras}

\usepackage{newtxtext,newtxmath}

\usepackage[T1]{fontenc}
\usepackage{ae,aecompl}
\usepackage{soul}
\usepackage{color}

\usepackage{graphicx}	
\usepackage{amsmath}	
\usepackage{amssymb}	
\usepackage{float}

\title[ELT Prototype of LGS wavefront sensor]{Prototype of a laser guide star wavefront sensor for the Extremely Large Telescope}
\author[M. Patti et al.]{
M. Patti$^{1,2}$\thanks{E-mail: mauro.patti@oabo.inaf.it },
M. Lombini$^{1}$, L. Schreiber$^{1}$, G. Bregoli$^{1}$, C. Arcidiacono$^{1}$, G. Cosentino$^{2}$,
\newauthor \mbox{} E. Diolaiti$^{1}$ and I. Foppiani$^{1}$ 
\\
$^{1}$INAF - Osservatorio Astronomico di Bologna, via Gobetti 93/3  40129 Bologna (Bo), Italy\\
$^{2}$UNIBO - Dipartimento di Fisica e Astronomia, via Gobetti 93/3  40129 Bologna (Bo), Italy\\
}

\date{Accepted 2018 March 16. Received 2018 February 09; in original form 2017 December 05}

\pubyear{2018}

\begin{document}
\label{firstpage}
\pagerange{\pageref{firstpage}--\pageref{lastpage}}
\maketitle

\begin{abstract}
The new class of large telescopes, as the future ELT, are designed to work with Laser Guide Star (LGS) tuned to a resonance of atmosphere sodium atoms. This wavefront sensing technique presents complex issues for an application to big telescopes due to many reasons mainly linked to the finite distance of the LGS, the launching angle, Tip-tilt indetermination and focus anisoplanatism. The implementation of a laboratory Prototype for LGS wavefront sensor (WFS) at the beginning of the phase study of MAORY, the Multi-conjugate Adaptive Optics RelaY for the ELT first light, has been indispensable to investigate specific mitigation strategies to the LGS WFS issues. This paper shows the test results of LGS WFS Prototype under different working conditions. The accuracy within which the LGS images are generated on the Shack-Hartmann (SH) WFS has been cross-checked with the MAORY simulation code. The experiments show the effect of noise on the centroiding precision, the impact of LGS image truncation on the wavefront sensing accuracy as well as the temporal evolution of sodium density profile and LGS image under-sampling.
\end{abstract}




\section{Introduction}
\begin{figure}
	\includegraphics[width=\columnwidth]{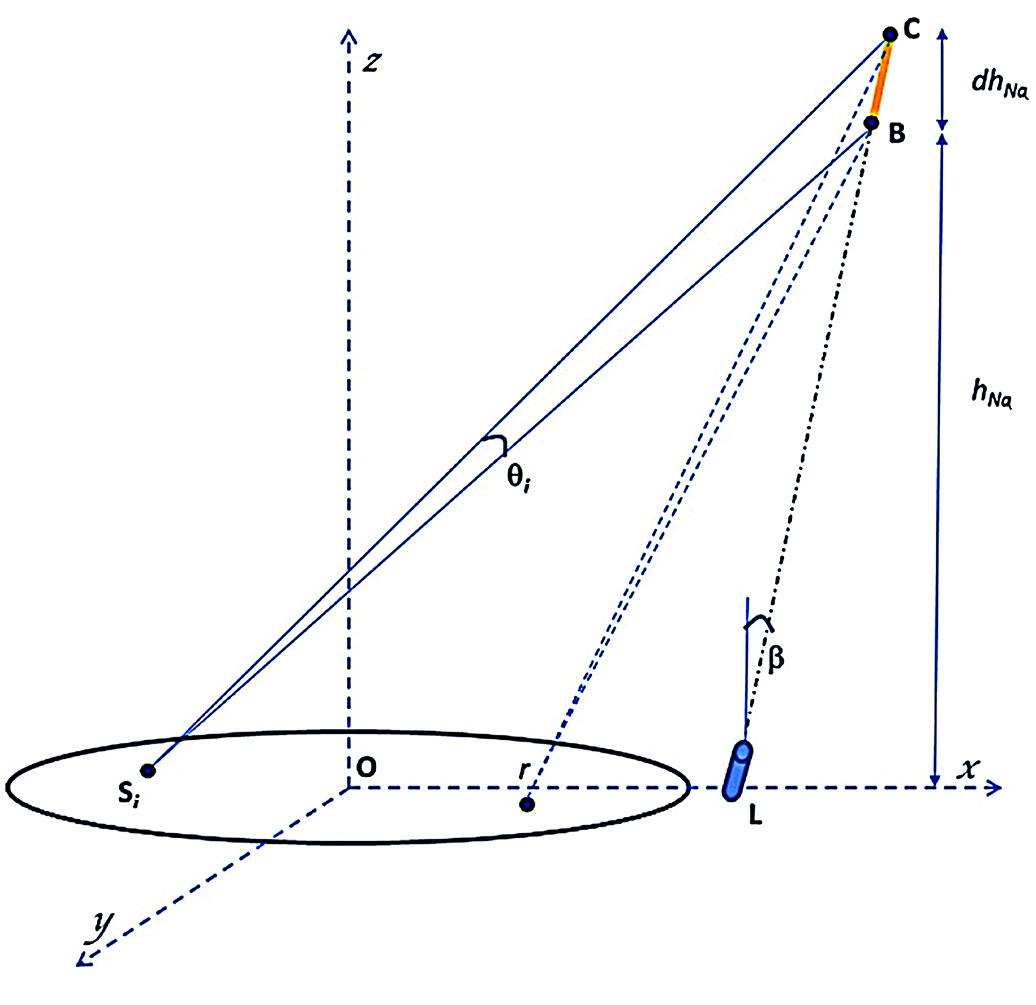}
    \caption{Geometrical representation of the LGS view from the SH sub-apertures.}
    \label{fig:1_f}
\end{figure}

\begin{figure}
	\includegraphics[width=\columnwidth]{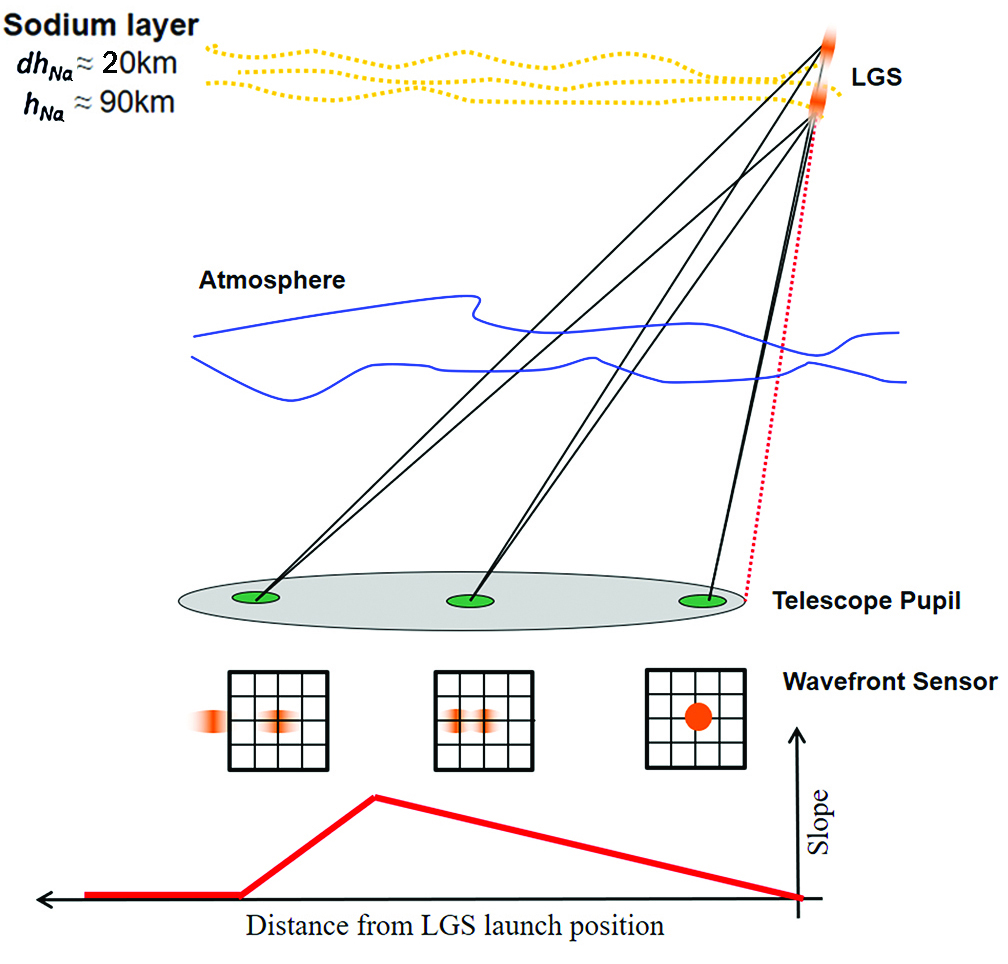}
	\caption{An example of bi-modal sodium density profile. Increasing the distance from the laser launcher position, the sub-aperture sees two distinct images. There is a certain position where the finite sub-aperture FoV produces a truncation of the second spot and creates a discontinuity in the slope measurements.}
	    \label{fig:1_ff}
\end{figure}

Wavefront sensing assisted by LGSs \citep{foy1985feasibility} is considered essential for the Adaptive Optics (AO) systems of future ELTs to achieve the required performance with high sky coverage. On the ELT \citep{gilmozzi200842}, six sodium LGSs are planned. These are generated by projecting powerful laser beams (589nm wavelength) from the edge of the telescope aperture up to the natural sodium layer in the mesosphere at about 90 km height. The return light samples the turbulent atmosphere and is collected by a set of WFSs, which measure in real-time the wavefront perturbations due to the atmospheric turbulence. Despite being bright and deployable over the whole observable sky, LGSs do not measure reliably some low orders of wavefront perturbations, such as tip-tilt (associated to image motion) and defocus \citep{rigaut1992laser}. For this reason, LGS need to be supplemented by additional measurements performed on Natural Guide Star (NGS), which may be however very faint \citep{wizinowich2006wm}. 
MAORY \citep{diolaiti2010maory} contains an optical relay \citep{lombini2016optical}, which creates an image of the telescope focal plane (entrance optical interface of MAORY) for the science instrument located on the MAORY exit port. The LGS light is propagated through this relay up to a dichroic beam-splitter, which is located after the deformable mirrors in order to let the LGS WFSs operate in close loop regime. The dichroic lets the light of 6 LGSs, arranged on a circle of about 120 arcsec diameter, pass through and reflects science beam and NGS light. Behind the dichroic an objective creates the LGS image plane for the WFSs channel. The LGS WFS will be based on the Shack-Hartmann (SH) WFS concept \citep{platt2001history}.
The sodium layer typically spans 10-20 km in height. The LGS, as observed from a large aperture telescope, such as the 39m diameter ELT, looks elongated because of a geometric perspective effect: the elongation increases with the distance from the laser launcher position. The wavefront measurement accuracy is affected by the LGS extension. 
In Figure~\ref{fig:1_f}, \textbf{O} is the origin of the $xyz$ reference frame and it is located in the centre of the telescope pupil having diameter = $2r$. \textbf{B} and \textbf{C} define the extremities of the LGS region of interest (e.g. sodium density profile FWHM when a single Gaussian density profile is considered) and have coordinates in the $xyz$ space respectively: 
\begin{equation}
\begin{aligned}
&\textbf{B}=[h_{Na}\tan\beta_{x}+x_{L} \mbox{ }; h_{Na}\tan\beta_{y}+y_{L} \mbox{ }; h_{Na}] 
\\
&\textbf{C}=[(h_{Na}+dh_{Na})\tan\beta_{x}+x_{L} \mbox{ }; \\&\mbox{ }\mbox{ }\mbox{ }\mbox{ }\mbox{ }\mbox{ }\mbox{ }\mbox{ }(h_{Na}+dh_{Na})\tan\beta_{y}+y_{L} \mbox{ }; (h_{Na}+dh_{NA})]
\end{aligned}
\end{equation}
where the coordinates of the Laser Launching Facility are [$\ x_{L} \mbox{ }; y_{L} \mbox{ }; 0$ ] and $\beta$ is the Laser launching angle. 
A generic sub-aperture $Si$ having centre coordinates [$\ Si_{x} \mbox{ }; Si_{y} \mbox{ }; 0$ ] sees the LGS under an angle: 
\begin{equation}
\theta_{i}=\cos^{-1}\frac{\overrightarrow{S_{i} B}\mbox{ }\overrightarrow{S_{i} C}}{ \lVert\overrightarrow{S_{i} B}\rVert \mbox{ }\lVert\overrightarrow{S_{i} C}\rVert}
\end{equation}
Considering a circular aperture of diameter D = 38.542m and a SH-WFS having 80 sub-apertures across the diameter, the worst case in terms of elongation is represented by the sub-aperture at the opposite side of laser launcher position. An interesting effect arises in the LGS WFS from the combination of the finite WFS Field-of-View (FoV) and of the sodium profile features. Consider Figure~\ref{fig:1_ff} where the sodium density profile has an ideal bi-modal distribution consisting only of two distinct spots separated by a certain altitude range. At the laser launcher position, the closest LGS WFS sub-aperture sees a negligible perspective elongation and the two spots overlap each other. If the LGS WFS is focused on one of these two spots, increasing the distance from the laser launcher position, the sub-aperture sees two distinct images. 
The one in focus at the centre of the sub-aperture FoV and the other closer to the sub-aperture FoV edge. The effect is a linear centroid shift of the LGS image with the distance from the laser launcher. There is a certain position where the finite sub-aperture FoV produces a truncation of the second spot and creates a discontinuity in the centroid measurements (and hence the slopes) across the pupil. The result is a more complex wavefront error (WFE) than pure Tip-Tilt and defocus (\citet{diolaiti2012dual}). The coupled effect of LGS spot truncation and sodium profile temporal variations are spurious and variable wavefront modes, which are only seen by the LGS WFS. Their injection into the AO correction loop is detrimental to image quality and therefore an external reference, typically a WFS working on NGSs, is needed for keeping these modes under control. This external reference, independent from sodium layer issues, in MAORY is the so-called reference WFS provided by an additional channel of the NGS WFS \citep{schreiber2009laser}.

\section{Laboratory experiment}
The experimental work has been conducted by means of a laboratory prototype of a LGS WFS \citep{patti2015laboratory} developed at INAF-OAS. The prototype reproduces the expected conditions, in the ELT case, when measuring the wavefront of LGS by means of a SH-WFS.
The laboratory experiment has been designed to achieve two main objectives:
\begin{itemize}  
\item	Experimental verification of LGS WFS performance under representative conditions for an ELT;
\item	Supporting the design of the wavefront sensing system of the MAORY instrument for the ELT.
\end{itemize}
A simplified version of the prototype was successfully integrated and tested in 2010 (\citet{lombini2010prototype}). It was able to generate realistic WFS data, including LGS spot perspective elongation and sodium profile features. The prototype has been upgraded to improve the accuracy on the generation of the desired sodium layer and to simulate, not simultaneously, a multiple LGS launching system.
The prototype is based on a SH-WFS with $40\times40$ sub-apertures: this WFS order is representative of the ELT, considering that the LGS WFS on the ELT will have typically $80\times80$ sub-apertures.
\begin{figure}
	\includegraphics[width=\columnwidth]{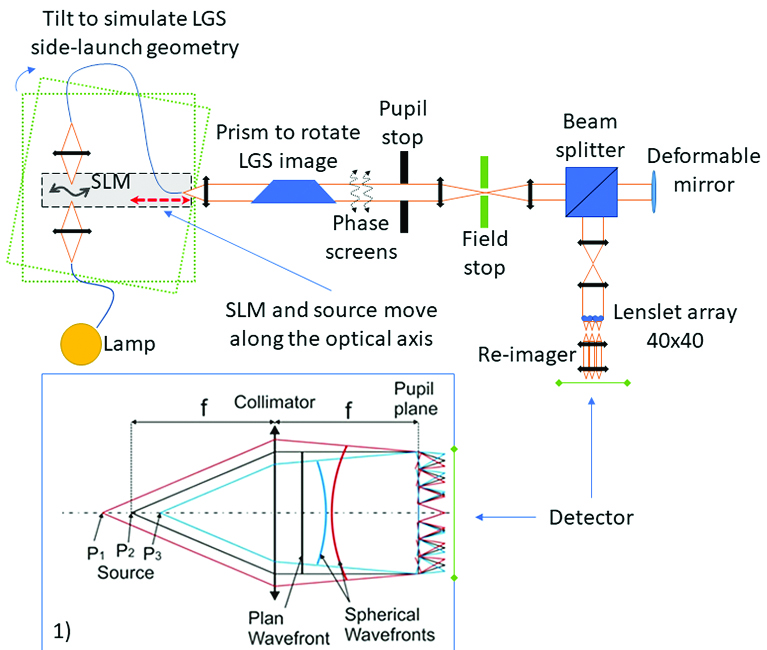}
    \caption{Prototype conceptual scheme. The axial motion of the source is a variable defocus that translates into a lateral motion of the spots in the WFS sub-apertures. The light intensity modulation is applied by a transmissive SLM. A tilt, respect to the optical axis, simulates the LGS side launch configuration. The prism, rotatating around the optical axis, changes the LGS launcher position around the pupil. Phase screens can be placed before the pupil stop. After the DM, conjugated to the pupil, a re-imager module is necessary to fit the lenslet array pitch with an integer and even number of detector pixels per sub-aperture.}
    \label{fig:2_f}
\end{figure}
The equivalent sodium layer extension which can be imaged by the prototype is about 20 km in the most elongated sub-apertures, corresponding to about 20 arcsec on the ELT. The conceptual scheme of the experimental setup is shown in Figure~\ref{fig:2_f}. The light source in the prototype is the output end of an optical fibre, which is fed by an intensity-modulated light pencil. The intensity modulation is applied by a remotely controlled spatial light modulator (SLM), a HOLOEYE LC 2002 Transmissive SLM. The SLM is placed between a polariser and an analyser in the so called amplitude modulation transmissive mode. This configuration permits the arbitrary choice of transmittance set through 256 values for each SLM pixel (0=no light to 255=full light) in order to simulate a realistic sodium density profile. The output end of the optical fibre and the SLM are both mounted on a motorised linear stage, while the input end of the optical fibre and the optics producing the light pencil feeding the fibre are fixed. The range of motion of the linear stage (about 10 mm) corresponds to the sodium layer extension which can be imaged on the camera. Along the linear stage motion, the output end of the fibre spans the full sodium layer extension while the input light pencil crosses different portions of the SLM surface. The axial motion of the source is a variable defocus that translates into a lateral motion of the spots in the WFS sub-apertures. The camera exposure time is set to a few seconds, corresponding to the time necessary for a full span through the sodium layer drawn on the SLM. A tilt of the linear stage travel axis with respect to the prototype optical axis, produces the typical elongation pattern, which is not radial, of a side-launch geometry of the LGS.
The light source module in the prototype includes a Dove prism mounted on a motorised rotation stage whose axis is co-aligned with the optical axis. By changing the position angle of the prism, it is possible to change configuration as if the LGS was launched from a different position around the edge of the telescope which is emulated by the prototype.
Atmospheric-like turbulence may be optionally generated by two plastic screens placed at the aperture stop of the prototype. The screens are mounted on X-Y linear stages to possibly apply temporal evolution.
A field stop is placed at an intermediate focal plane: its diameter can be manually adjusted to introduce different truncations on the LGS images produced by the WFS.
The prototype includes a low-order deformable mirror (ALPAO DM52 with 52 actuators). The purpose of the DM is to introduce low-order and static wavefront aberrations for the test described in subsection~\ref{subsec:under-sampling}. In the remaining test cases, there is no voltage applied to the actuators and the DM surface can be approximated to a flat rigid mirror.
The core of the Shack-Hartmann WFS is a lenslet array, characterised by square geometry, pitch size 300$\mu$m, focal length 3.82 mm.
The array of Shack-Hartmann images is recorded by a scientific-grade CCD camera (pixel size 13$\mu$m, $1k\times1k$ pixels). Each WFS sub-aperture is mapped onto $24\times24$ pixels of the CCD camera. A re-imager module is necessary to fit the lenslet array pitch with an integer and even number of detector pixels per sub-aperture.
The LGS spots are sampled by 2 pixels per FWHM along the non-elongated axis. Analogue on-chip binning or numerical binning may be used to reduce the sampling to only 1 pixel per FWHM and emulate severe under-sampling conditions.
All the prototype functions are controlled by custom control software coded in C/C++. The software controls all devices (motorised stages, SLM, DM, camera) and manages the prototype configuration setup and the acquisition cycles. Data reduction is performed by specific software coded in high-level language IDL\textsuperscript{\textregistered}.
\subsection{Common test conditions, limits and calibrations}
\label{subsec:Common} 
The Prototype was integrated under a controlled room temperature to reduce thermal fluctuations that effect the system performance. Fixing the temperature at$\ 23^oC$, the air conditioning system generates a periodic variation within$\ 1^oC$. This introduces a systematic error in the measures that is translated in global X-Y centroids offset in function of the temperature. The effect is a loss of precision, i.e. a dispersion of a set of WFS images. In terms of RMS WFE, $\approx$100nm of differential Tip and Tilt terms are introduced while the root sum of squared higher order terms is below 12nm.
Every test was performed in a regime of Signal-to-Noise Ratio (SNR) where the number of photons per sub-aperture ($n$) stay within 500 and 5000. The laboratory CCD camera has a RON of $\ 15e^-$/pixel. In a CCD RON-dominated:
\begin{equation}
SNR = \frac{n}{\sqrt{n+RON^2}}
\end{equation}
Scaling the SNR to a typical AO CCD RON of$\ 4e^-$/pixel, the equivalent number of detected photon per sub-aperture stay within 360 and 4800. 
In SH WFS, at linear regime, centroid coordinates are a direct measure of the local wavefront slope. The Center of Gravity (CoG) is a simple algorithm to estimate the spot position but it is very sensitive to noise. A solution to reduce the noise is to apply a threshold.
The threshold value $T_{th}$ is determined as follows:
\begin{equation}
T_{th} = T_{\%}\mbox{ }max(I)
\end{equation}
Where $\ T_{\%}$ is the percentage relative to the maximum intensity ($\ max(I)$). The threshold modifies the intensity distribution as follows:
\begin{equation}
I_{t} =\begin{cases}I - T_{th}  & 	\mbox{for }\mbox{ }   I \ge T_{th}  \\ 0   & \mbox{for }	\mbox{ }    I < T_{th} \end{cases}
\end{equation}
The CoG with a threshold reduces the noise but is still sensitive to the LGS density profile shape and temporal variation. A solution to mitigate these effects is to use the Weighted Centre of Gravity (WCoG) \citep{fusco2004optimization} algorithm which uses a ``reference image'' to evaluate the spot position and it is less sensitive to noise and LGS image intensity variations. 
The ``reference image'' can be assumed equal to the mean LGS image inside each sub-aperture within a given timescale\citep{schreiber2009laser}. The reference acts as a weighted function which reduces the noise effects but introduces an error on the centroid estimation that is proportional to the distance of the actual centroid from the centre of the weighting function. To compensate this error, a calibration curve is empirically derived in each sub-aperture\citep{schreiber2010laser}.
The prototype experiments didn't consider the temporal aspects of atmospheric turbulence and the LGS image, inside each sub-aperture, has a fixed position. This condition makes negligible any biasing effects due to the weighting function which are not considered. The experimental results, described in this paper, refer only to WFS performance using WCoG centroiding.
Centroid measures were translated in Optical-Path-Difference (OPD) as follows:
\begin{equation}
\begin{split}
 OPD_{(x)} = \frac{d}{f}\ \Delta x \\ OPD_{(y)} = \frac{d}{f}\ \Delta y 
\end{split}
\end{equation}
Where $d$ is the lenslet diameter and $f$ its focal length.  $\Delta x$ and $\Delta y$ are centroid coordinates respect to the sub-aperture centre.
Test results that consider the distribution of sub-aperture OPD are always corrected for the median RMS offset introduced by temperature variations. 
Test results that consider the distribution of RMS WFE are always corrected for prototype static aberrations that are measured by taking as reference a top-hat sodium profile that does not extend outside the field stop. These corrections on OPD and WFE are a kind of bias subtraction in the resulting data.
To retrieve a wavefront from centroid measures, the followed approach is a modal reconstruction method with Zernike polynomials \citep{noll1976zernike}. The numbers of modes used to fit the wavefront aberrations are 252 Zernike. Further increment on the number of modes results negligible in terms of fit residuals which are within numerical accuracy. Every time a total RMS WFE is computed, Tip-Tilt and defocus are excluded ($Z_{2}$ to $Z_{4}$). MAORY is designed to work with a reference WFS based on NGSs to monitor low-order wavefront aberrations. Thus, some tests also considered the total RMS WFE for Zernike modes above 54 ($\ Z_{54}$) as a relevant information to support the definition of the requirements of the NGS wavefront sensing sub-system.
The SLM displays a chosen sodium density profile on a grayscale from 0 to 255 levels that is not linear. Any chosen sodium density profile has to be calibrated by the SLM curve response \citep{patti2016accurate}.
The Prototype light source, a multimode optical fibre, delivers an output beam whose intensity distribution is approximately Gaussian in shape, involving a gradual drop of the pupil image intensity at its borders. The multimode optical fibre allows the propagation of different light modes whose superposition, at the output end of the fibre, generates lighting inhomogeneities at the pupil stop \citep{patti2016accurate}.
Simulated LGS launcher position is on the edge of telescope primary mirror and each sample of images does not consider the temporal aspects of atmospheric turbulence. Given a static sodium density profile, the smallest sample of images, used for the test statistic, contains 100 images. Data reduction takes care of background subtraction, hot pixels, bad lines and cosmic rays. Corrupted pixels values were replaced by the mean of adjacent pixels.
\section{Results}
Extensive test campaigns have been carried out with the experimental setup. Relevant results are shown in the following subsections.
\subsection{Simulation code verification}
\label{subsec:End-to-end} 
The MAORY simulation code has been designed to accurately model all the aspects of the LGS image in the SH-WFS sub-apertures and its temporal variation. An accurate description of the code is reported in \citet{arcidiacono2014end} and \citet{arcidiacono2016numerical}. 
\begin{figure}
\centering
	\includegraphics[width=\columnwidth]{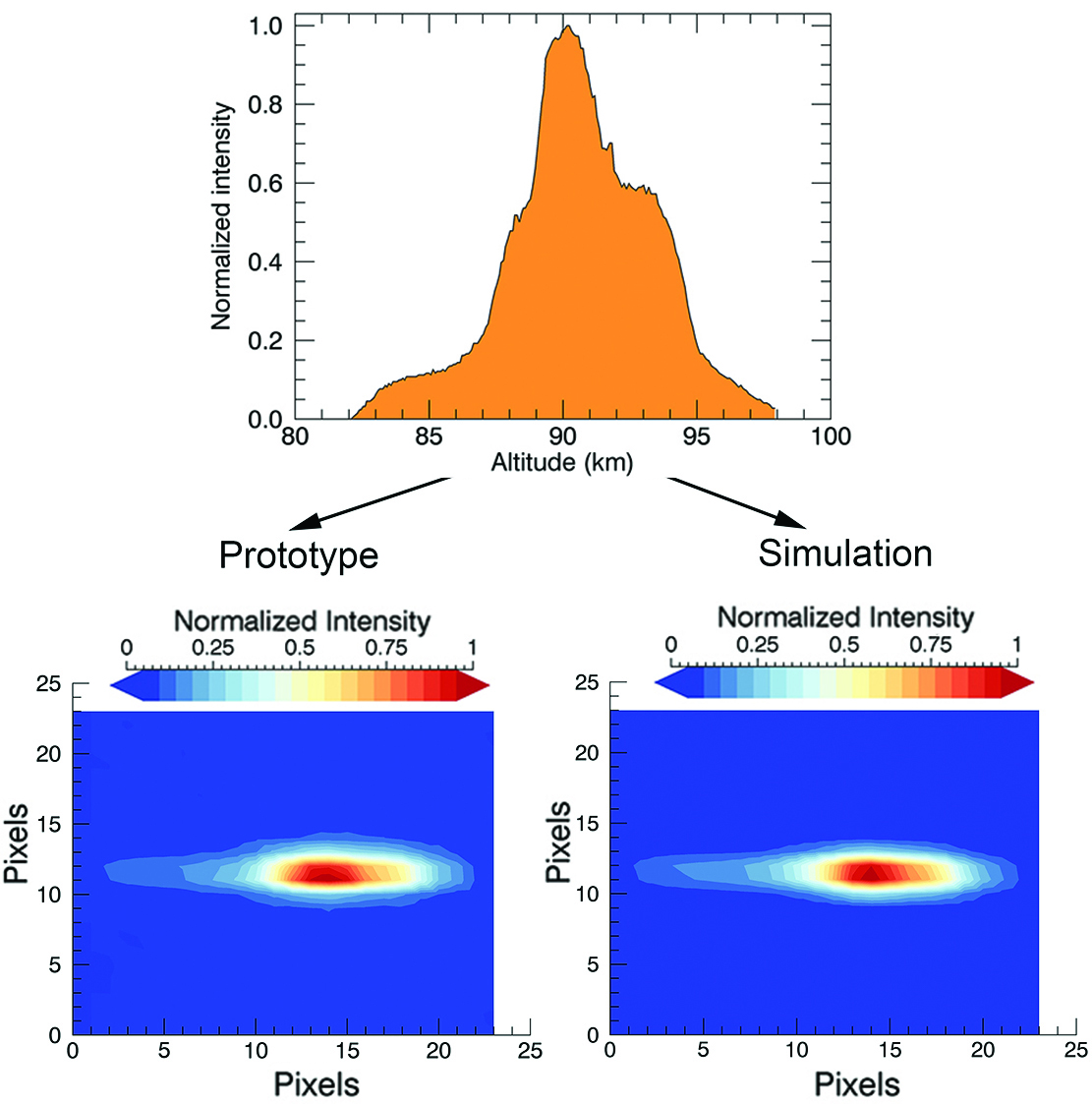}
    \caption{The sodium density profile spans about 20km in altitude. Simulation and prototype data of the most elongated LGS image, at the SH WFS, are shown.}
    \label{fig:3_f}
\end{figure}
The code is also able to simulate a transverse sodium density profile that, in a multiple LGS system, could lead to differential effects among the LGSs. The laboratory prototype has been essential to support the development of the simulation code. Given a sodium density profile as input, the verification on the accuracy with which the simulation code translates the input in LGS images at the WFS focal plane, has been done by the prototype data. \citet{patti2016accurate} reports a comparison between simulation and experimental data. A real sodium profiles from Lick Observatory \citep{thomas2010analysis} has been injected into the prototype and, for comparison, into the simulation code. In the simulation, any source of noise (i.e. photon noise, detector noise, background light, etc.) has been excluded as well as the atmospheric turbulence. During the prototype image acquisition, a very high SNR regime has been considered so that the images are almost noise free. Figure~\ref{fig:3_f}, as reported in \citet{patti2016accurate}, is a visual comparison of the sodium density profile as seen by the most elongated sub-aperture. As parameter for the comparison, we used the linear Pearson correlation coefficient which turned out to be equal to 0.97 implying a very high correlation between the two images.
\begin{figure}
\centering
	\includegraphics[width=\columnwidth]{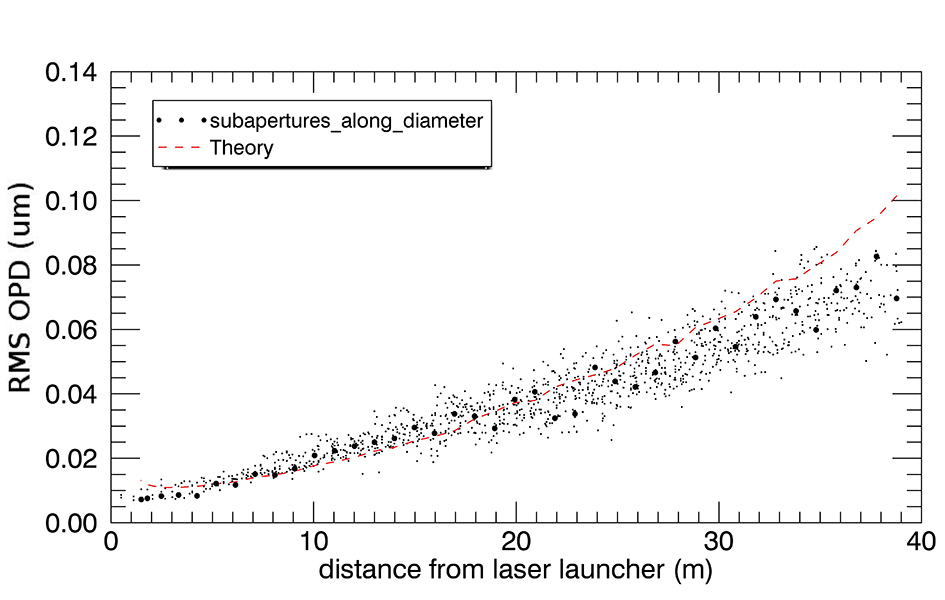}	
    \caption{Measurements performed by WCoG algorithm with a Gussian sodium density profile. The RMS OPD values of each sub-aperture are the black dots. The theoretical behavior in function of the distance from laser launcher is the dashed line. Bigger black dots are RMS OPD values of the sub-apertures where the LGS elongation is parallel to the lenslet side.}
    \label{fig:4_f}
\end{figure}
\begin{figure}
\centering
\includegraphics[width=0.85\columnwidth]{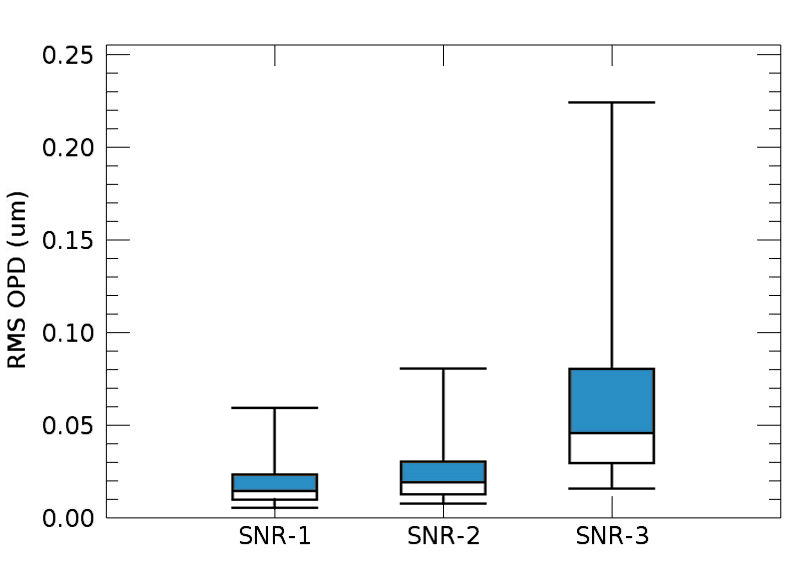}
\caption{RMS OPD distributions for three levels of SNR. Each box plot is the distribution of RMS OPD values of the sub-apertures. Number of photons per sub-aperture is $\approx$1800 for SNR-3, $\approx$2700 for SNR-2 and $\approx$4700 for SNR-1. Box plots show minimum and maximum values and quartiles.}
 \label{fig:4_ff}
 \end{figure}
\subsection{Wavefront sensor measurement error}
\label{subsec:WFE} 
The measurement error at WFS sub-aperture level has been studied for different conditions of SNR, sodium profile and centroid algorithm. In all the cases very good match has been found between experimental results and theoretical expectations derived from numerical simulations. In this subsection, the experimental results refer to a Gaussian sodium profile for three conditions of SNR.
Any centroid algorithms performance is mainly affected by the following sources of error:
\begin{itemize}
\item Photon noise which follows a Poisson distribution and starts to be significant when the light intensity is low.
\item Background noise which includes hot pixels, bad lines and cosmic rays.
\item Sampling error which is related to the detector pixels size.
\item Fixed pattern noise which is related to the digitization of light intensity by pixels.
\item Sidelobes of the spot irradiance distribution which depends not only on diffraction but also on optical surface imperfection (i.e. scratches, dig, bubbles). If the sidelobes symmetry was broken within the centroid searching area, it could cause centroid errors.
\end{itemize}
In typical LGS images delivered by the prototype is not possible to distinguish these error sources one from the other. For these reasons and many other aspects \citep{patti2016accurate}, the simulation and prototype data will always be different.
Figure~\ref{fig:4_f} shows the prototype WFS measurement errors in terms of RMS OPDs compared to the theoretical behavior \citep{schreiber2009laser}. Given a sample of images, the RMS OPD of each sub-aperture is the RMS position difference of centroid measurements with respect to the mean.
To assess the performances of WGoC for different levels of SNR, the mean photon counts per sub-aperture is considered as indicator of noise and the distribution of RMS OPD values of the sub-apertures is the performance measure. Figure~\ref{fig:4_ff} considers three levels of SNR increasing from SNR-3 to SNR-1. The equivalent number of photon counts per sub-aperture is $\approx$1800 for SNR-3, $\approx$2700 for SNR-2 and $\approx$4700 for SNR-1. Considering a sub-pupil of $0.5m$ and the LGS WFS operating at $700Hz$, these numbers are representative of expected LGS return flux per sub-aperture at ELT site \citep{madec2016adaptive} \citep{calia2016lgs}. The three distributions are RMS OPD values of $40\times40$ sub-apertures. As expected, more photon counts per sub-aperture means a better centroiding accuracy. 
 \begin{figure}
\centering
	\includegraphics[width=0.4\textwidth]{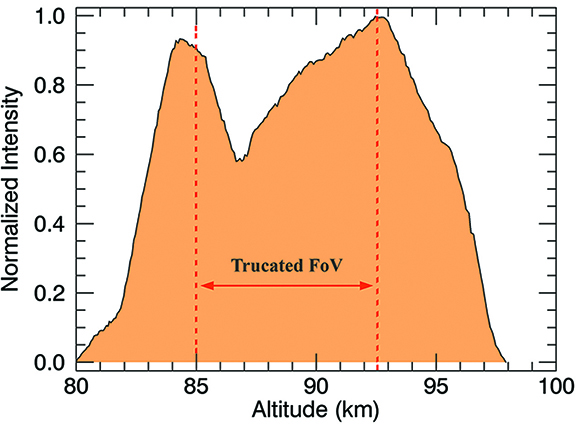}
    \caption{Realistic sodium density profile used to evaluate the effect of LGS image truncation. Altitude range (20km) refers to the maximum sub-aperture FoV (about 20 arcsec).}
    \label{fig:5_f}
\end{figure}
\subsection{Effect of image truncation }
\label{subsec:truncation} 
\begin{figure}
\centering
	\includegraphics[width=0.35\textwidth]{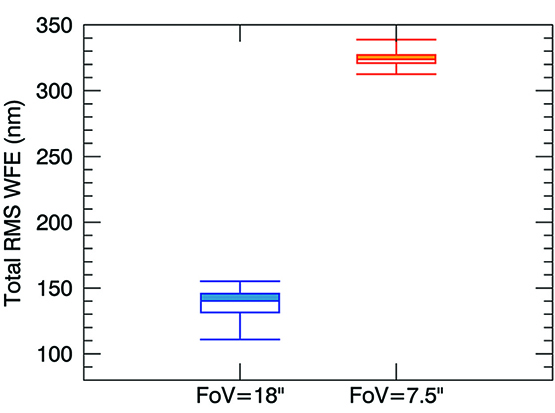}
	\includegraphics[width=0.35\textwidth]{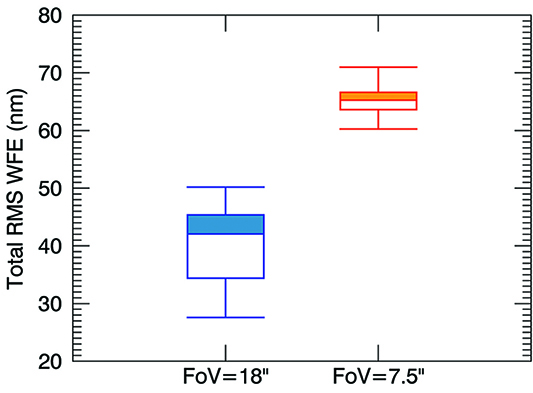}
    \caption{WCoG algorithm results described in subsection~\ref{subsec:truncation}. Upper plot: Total RMS WFE above $Z_{4}$; Lower plot: Total RMS WFE above $Z_{54}$. Given a sample of WFS images, these are the distributions of the total RMS WFE (in nanometres) introduced by the static sodium density profile of Figure~\ref{fig:5_f} for two different FoV. Box plots show minimum and maximum values and quartiles.}
    \label{fig:6_f}
\end{figure}

\begin{figure}
\centering
	\includegraphics[width=0.35\textwidth]{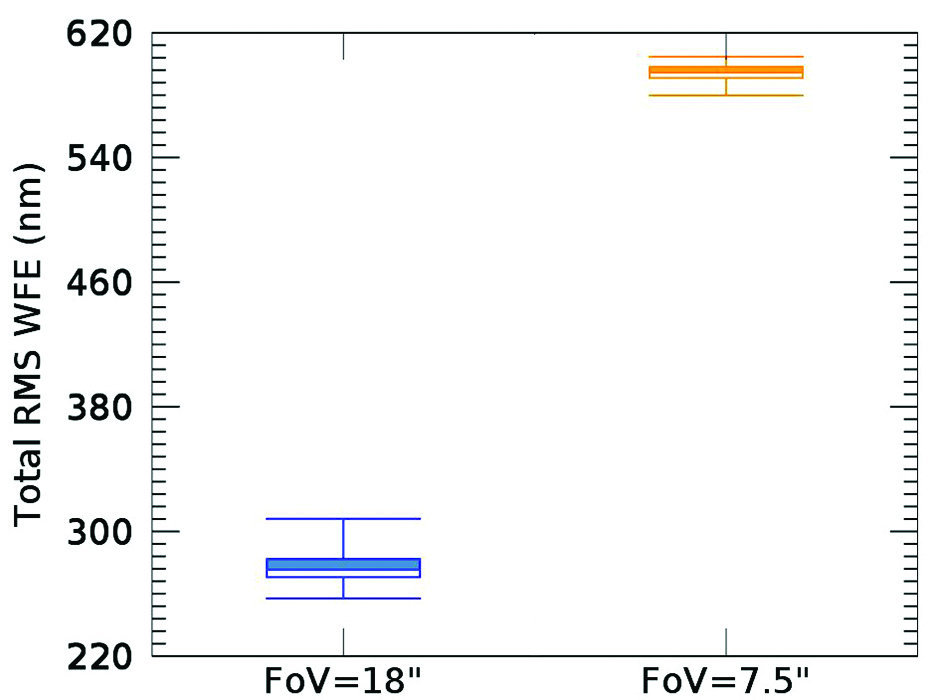}
	\includegraphics[width=0.35\textwidth]{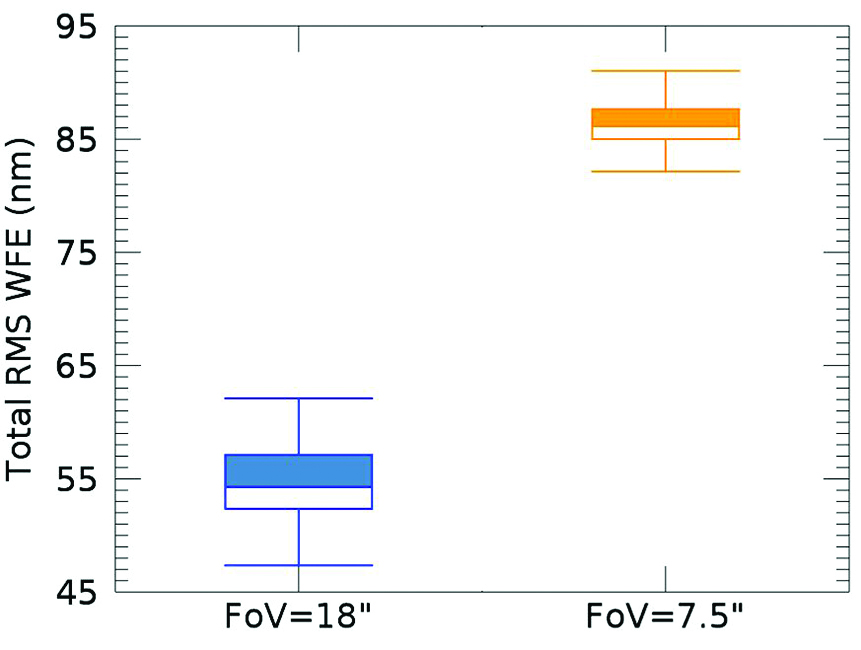}
    \caption{Same of Figure~\ref{fig:6_f} but without calibration for prototype static aberrations.}
    \label{fig:7_f}
\end{figure}

Truncation effects make the WFS detect additional spurious aberrations, especially affecting the first few tens of Zernike modes, which would require a kind of on-line calibration of the LGS WFS measurements by additional reference measurements (e.g. to be performed on NGS). 
For comparison, we selected a sodium density profile entirely contained in the maximum sub-aperture FoV which is varied by means of a field stop. The following results are related to the sodium density profile shown on Figure~\ref{fig:5_f}. The image truncation is symmetric respect the two density peaks. Figure~\ref{fig:6_f} shows the difference in terms of total RMS WFE for sub-aperture FoV of 18 arcsec and 7.5 arcsec. Given a sample of WFS images at the same test conditions, each box plot is the RMS WFE distribution of the sample. These results are not influenced by data calibrations described in subsection~\ref{subsec:Common} since, as already mentioned, the calibration procedure is a kind of bias sustraction of resulting data as shown in Figure~\ref{fig:7_f}. The measured wavefronts, with and without image truncation, are not corrected for prototype static aberrations and lead to the same conclusions. The LGS truncation introduces hundreds of nanometres of low-order WFE and it is less significant for modes above $Z_{54}$.
It has to be considered a loss of photons of about 10\% on the entire pupil due to the field stop. The effect of different SNR combined to a fixed level of truncation is the subject of next session.
\subsection{Truncation vs SNR}
\label{subsec:truncation vs SNR} 
\begin{figure}
\centering
	\includegraphics[width=0.4\textwidth]{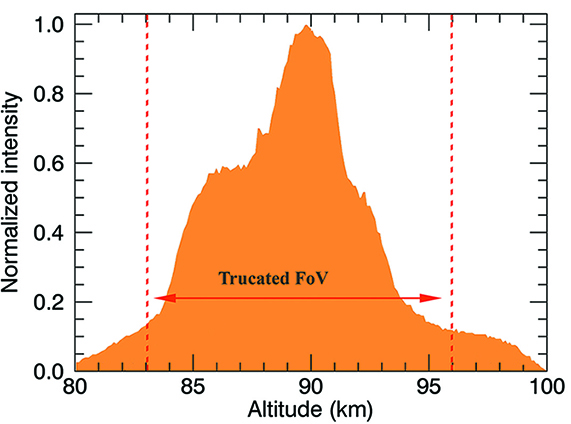}
    \caption{Realistic sodium density profile used to evaluate the effect of LGS image truncation vs SNR. Altitude range (20Km) refers to the maximum sub-aperture FoV (about 20 arcsec).}
    \label{fig:8_f}
\end{figure}

This subsection evaluates the effect of LGS image truncation compared to different levels of SNR. The selected sodium density profile is entirely contained in the maximum sub-aperture FoV which is varied by means of a field stop. 
The following results are related to the sodium density profile showed in Figure~\ref{fig:8_f}. The image truncation regards the low end of sodium density distribution. The number of photons per sub-aperture are regulated by changing the source light intensity. Two levels of SNR have been set when there is no truncation while a fixed SNR, that falls between the other two, has been set when truncation occurs.
\begin{figure}
\centering
	\includegraphics[width=0.35\textwidth]{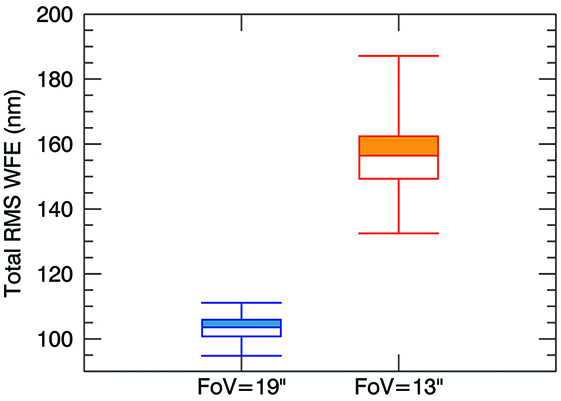}
	\includegraphics[width=0.35\textwidth]{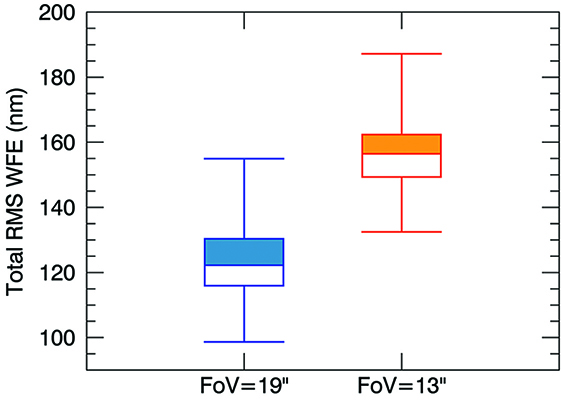}
    \caption{WCoG algorithm results described in subsection~\ref{subsec:truncation vs SNR}. Statistical distribution of the total RMS WFE (in nanometres) introduced by the static sodium density profile of Figure~\ref{fig:8_f} for two different FoV. Truncated LGS image is the orange boxplot. Upper plot: Comparison with higher level of SNR of not truncated case (blue); Lower plot: Comparison with lower level of SNR of not truncated case (blue). The box has lines at the lower-quartile, median, and upper-quartile values.}
    \label{fig:9_f}
\end{figure}
Figure~\ref{fig:9_f} shows the difference in terms of total RMS WFE for a sub-aperture FoV of 19 arcsec and 13 arcsec. Given a sample of WFS images at the same test conditions, each box plot is the RMS WFE distribution of the sample. Blue boxes refer to no-truncated profile whose SNR decreases going from the upper plot to the lower plot. Orange boxes refer to truncated profile whose SNR is fixed. The effect of reduced SNR translates to greater RMS WFE as well as a wider distribution due to noise. However, the medians of RMS WFE distributions of not truncated cases (blue), are lower than the truncated case (orange).

\subsection{Effect of image under-sampling}
\label{subsec:under-sampling} 
To avoid LGS truncation, for a given number of pixels, the sub-aperture FoV may be increased by under-sampling. However, in this case the WFS loses linearity. Two different approaches are under investigation: 
\begin{itemize}
\item	To calibrate the gain of the centroid algorithm for non-linearity effects introducing a known periodic tilt signal on both axes via a LGS WFS jitter compensation mirrors; 
\item	To introduce a blur in the LGS image in order to re-cover Nyquist sampling in the non-elongated axis, with a negligible effect on the elongated axis. 
\end{itemize}
\begin{figure}
\centering
	\includegraphics[width=0.45\textwidth]{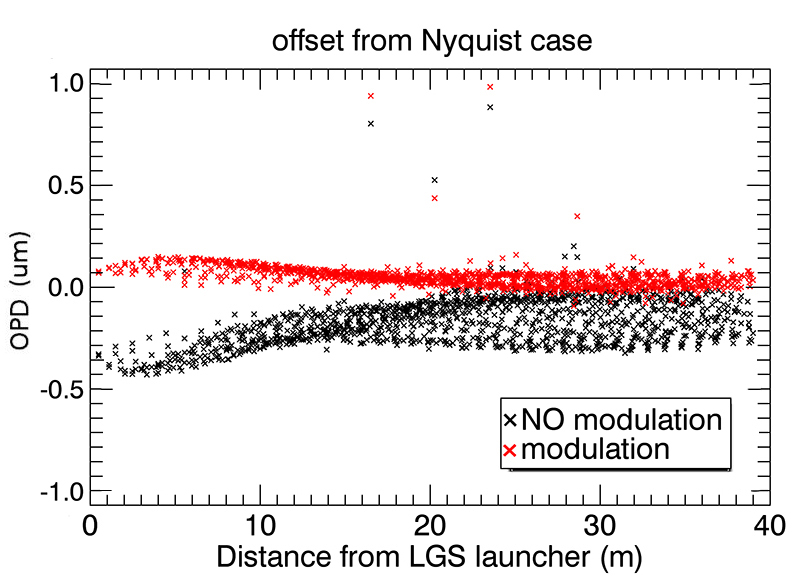}
	\includegraphics[width=0.5\textwidth]{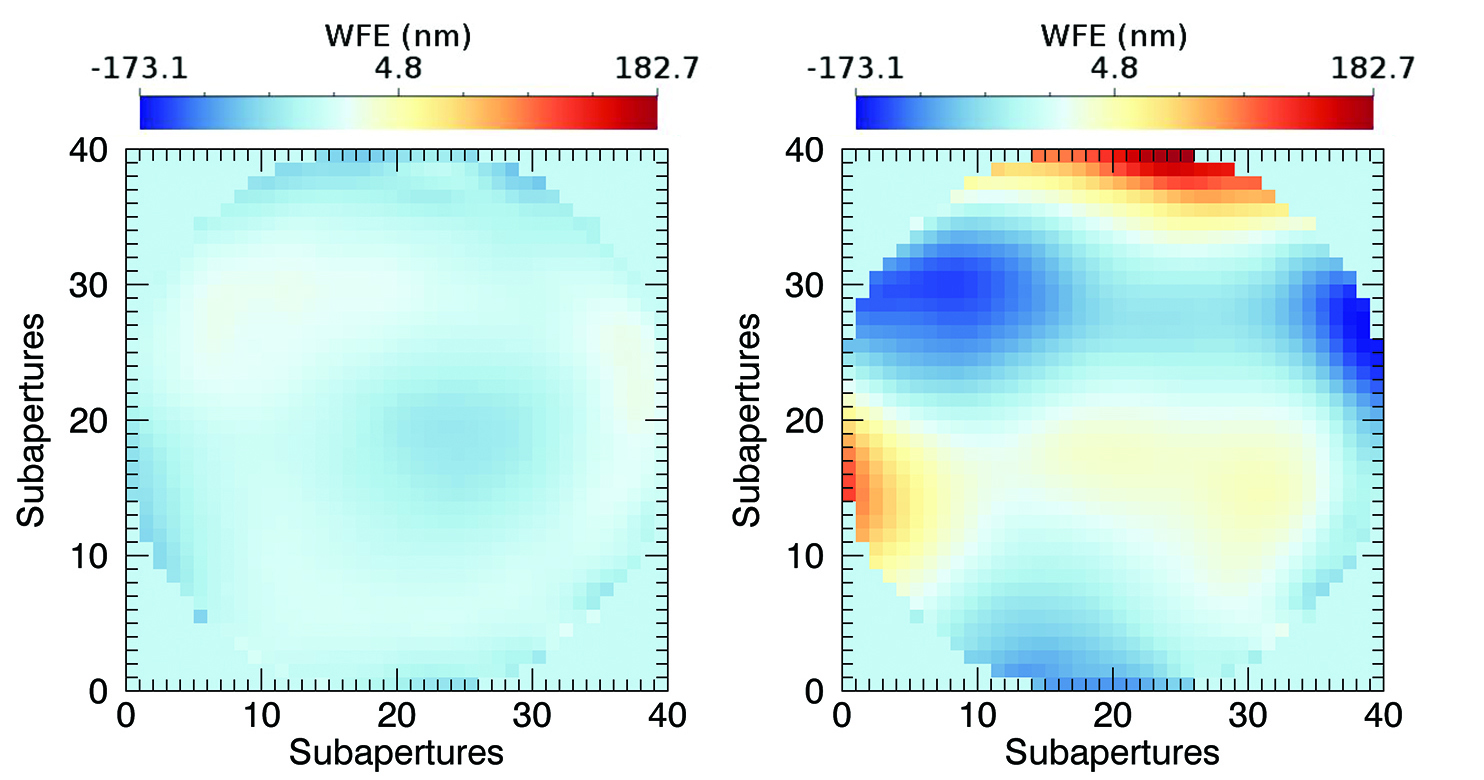}
    \caption{Test results described in subsection~\ref{subsec:under-sampling}. The sodium profile of Figure~\ref{fig:8_f} has been used. Upper plot: Centroid offset on under-sampled data, with respect to centroid measurements obtained on Nyquist-sampled data, as a function of sub-aperture distance from the laser launching position. Red points are OPDs after calibration for under-sampling; black points are OPDs without calibration. Lower plot: Residual wavefront map (nm) due to centroid offsets with calibration (left) and without calibration (right).}
    \label{fig:10_f}
\end{figure}
The first approach is explained in this section. Under-sampled
WFS data have been produced by numerical binning of Nyquistsampled
data. To mitigate the effect of under-sampling, the implemented
gain centroid calibration procedure is based on spot modulation
(i.e. dithering) by numerically shifting the Nyquist-sampled
LGS images by an integer number of pixels. The dithering is done
before applying the 2$\times$2 numerical binning to avoid errors due to
numerical interpolation ($\pm1$ pixel). The modulated images are then
binned and a calibration curve is derived for each sub-aperture. This
procedure simulates the presence of a jitter mirror in each LGS WFS
channel, which introduces a known sub-pixel shift in the LGS image
for calibration. The calibration curve links the detected centroids to
the true amount of image shift. The calibrated centroids are defined
as:
\begin{equation}
C(x,y) = \begin{matrix} \sum_{n} a_{n}(x,y)\mbox{ }C_{o}^n(x,y) \end{matrix}
\end{equation}
where$\ a_{n}(x,y)$ are the coefficients of the calibration curve,$\ C_{o}^{n}(x,y)$ are the $x$ and $y$ measured centroids and $n$ is the polynomial degree used to fit the curve. As a first approximation,  first degree polynomial was used. 
The sodium profile of Figure~\ref{fig:8_f}has been used
without truncation to disentangle non-linearity effects due to undersampling
and non-linearity effects due to LGS image truncation.
The test was conducted for very high SNR to avoid centroid errors
due to noise. Through the DM integrated in the prototype, a certain
amount of low-order aberrations can be introduced to shift the LGS
spots within the sub-apertures and therefore, force the WFS to work
in a non-linear regime (for the under-sampling condition). A single,
noise-free and Nyquist-sampled LGS image has been acquired with
the SH-WFS. The measured spot centroids are used as reference.
Applying the 2$\times$2 numerical binning on the image, new centroid
coordinates are detected and their offset from the Nyquist case is the
performance measure. After the calibration, the spot centroid offset
due to under-sampling was again evaluated using the same reference.
To retrieve the wavefront from the centroid measurements,
the approach followed is a modal reconstruction with Zernike polynomials.
Offsets in terms of spot centroids translate into residual
WFE.

The results, in terms of OPD, are shown in Figure~\ref{fig:10_f}. Each dot is a single sub-aperture offset from the Nyquist case. They have been computed by subtracting the centroid values in the case of 2$\times$2 image binning from the corresponding centroid values in the case of Nyquist sampling and summing in quadrature the x and y OPDs. Black dots correspond to OPDs offsets due to the absence of a gain calibration procedure, while red dots correspond to OPD offsets when the calibration procedure is applied. It is clear that sub-apertures near the laser launcher position strongly suffer the under-sampling effect because in that region of the pupil, where there is a moderate LGS elongation, the spots are severely under-sampled in both axes. The wavefront maps on Figure~\ref{fig:10_f} refer to residual aberrations ($Z_{2}$ to $Z_{4}$ excluded) introduced by the effect of under-sampling in the cases where a centroid calibration procedure does (left panel) and does not (right panel) occur. 
A centroid calibration, based on spot modulation, mitigates the effect of under-sampling reducing the residual RMS WFE from $\approx$103nm to $\approx$26nm. This means that the deviation from a Nyquist detector is reduced by a factor of $\approx$4 when using spot modulation.
Of course, these results were obtained on a scaled model ($40\times40$ sub-apertures instead of $80\times80$) and under simplified experimental conditions.
\subsection{Effect of varying the sodium density profile}
\label{subsec:profile variation} 
In this section, the effect of a variable sodium density profile is evaluated in terms of rms WFE. 11 sodium density profiles were selected from 1 h of night observations. We chose extreme cases with highly structured density profiles as shown in Figure~\ref{fig:11_f}. The
profiles are not entirely contained in the maximum sub-aperture
FoV, so spot truncation unavoidably occurs. This condition permits
us to test a device that monitors the sodium temporal evolution and,
by talking to the AO system, changes the LGS reference altitude
according to the number of significant sodium density structures.
By keeping the maximum number of major density peaks inside the
sub-aperture FoV, such a tool could:
\begin{figure}
	\includegraphics[width=0.3\textwidth]{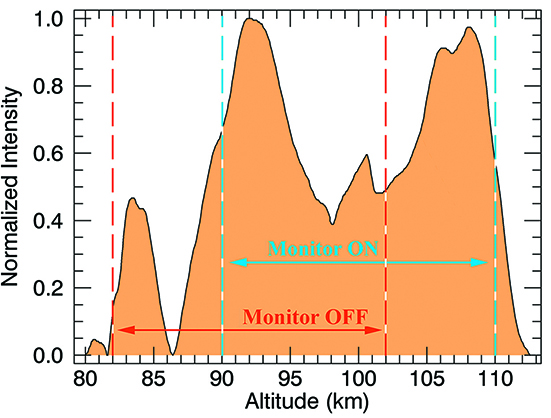}
	\includegraphics[width=0.17\textwidth]{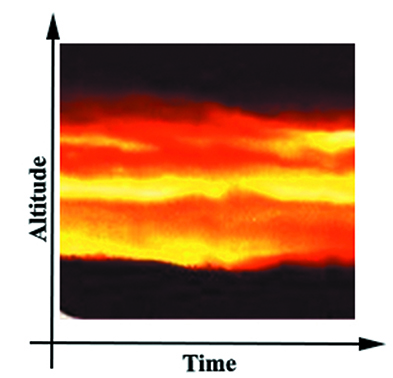}
    \caption{Sequence of sodium density profile in function of Time and Altitude (right). One of the profile extracted from the sequence and used for test measures (left). Herein is showed an example of how the maximum sub-aperture FoV changes in the case we are monitoring (blue) or not monitoring (red) the complete sodium density distribution.}
    \label{fig:11_f}
\end{figure}
\begin{figure}
\centering
	\includegraphics[width=0.4\textwidth]{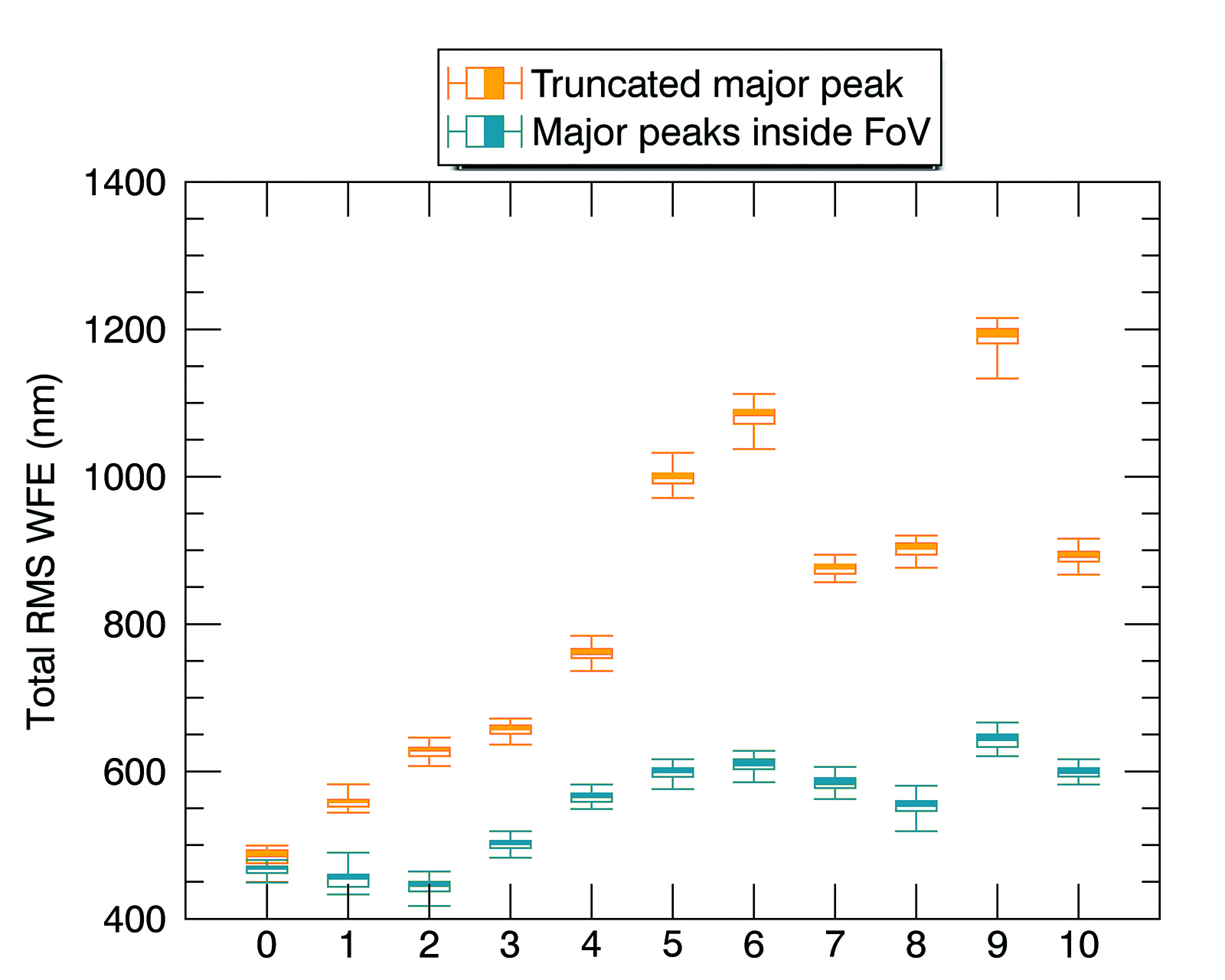}
	\includegraphics[width=0.4\textwidth]{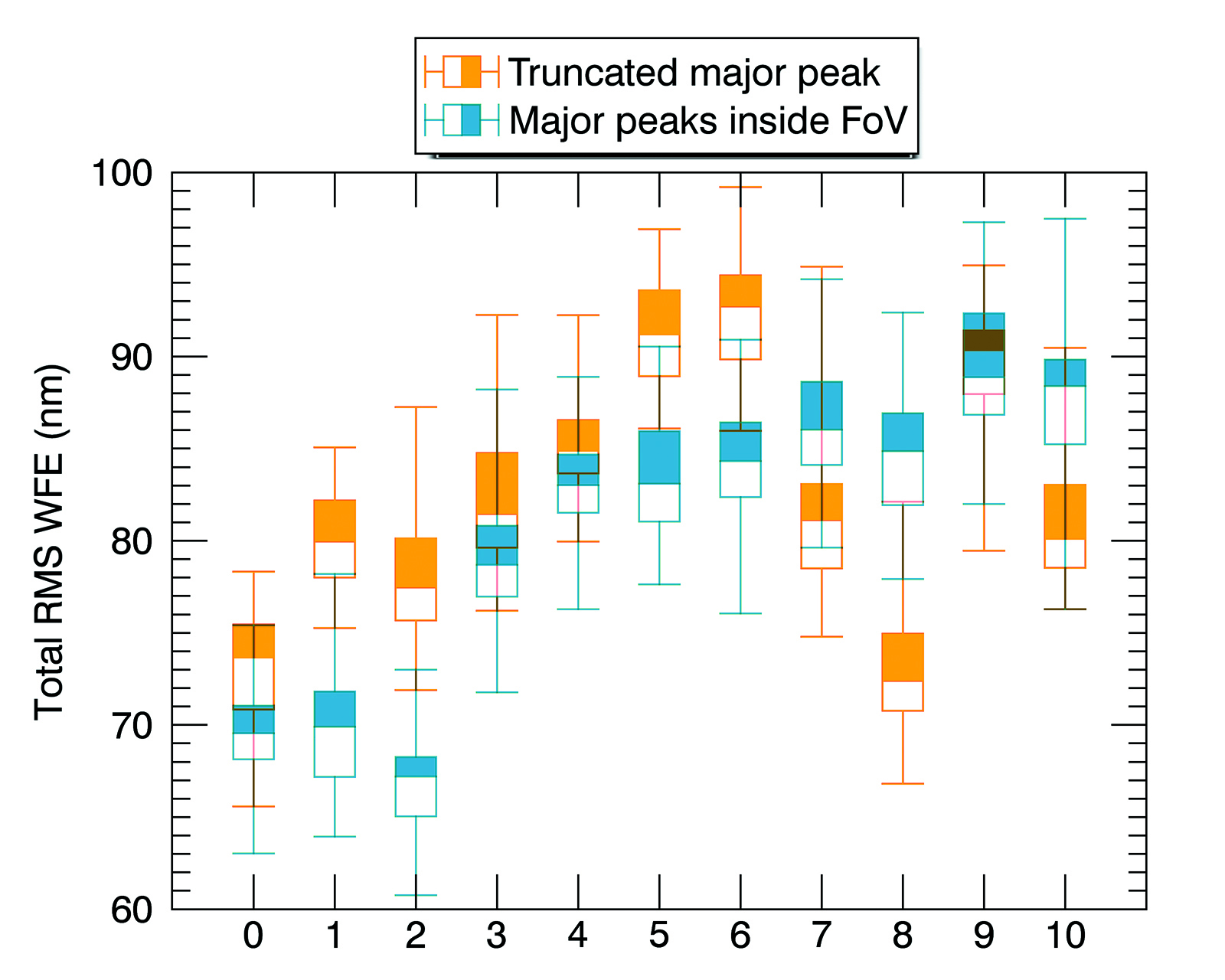}
    \caption{WCoG algorithm results described in subsection~\ref{subsec:profile variation}. Upper plot: Total RMS WFE above $Z_{4}$; Lower plot: Total RMS WFE above $Z_{54}$. Each boxplot refers to a static sodium density profile that is different for each point of the plot. In blue: Applied principle of sodium density monitor; In orange: Not applied principle of sodium density monitor that causes a large spot truncation.}
    \label{fig:12_f}
\end{figure}
\begin{itemize}
\item	prevent spurious aberrations, especially if a large spot truncation is foreseen.
\item	provide consistent statistic of sodium density profiles that might correlate with AO system performance.
\end{itemize}
Figure~\ref{fig:12_f} shows the total RMS WFE related to 11 sodium density profiles in the cases where, in principle, we can (blue) and cannot (orange) use a real-time density monitor tool. To understand the working condition of the tests, we must refer to Figure~\ref{fig:11_f}. If we could know the complete sodium density profile, we would change the LGS reference altitude in order to reduce the truncation of density major peaks. This translates to an overall lower RMS WFE but the results are comparable considering Zernike modes above $Z_{54}$. The profile variation still affects these high-order modes but it is less significant. The temporal evolution of sodium density profile introduces hundreds of nanometres of WFE and it is comparable to the impact of image truncation on the wavefront sensing accuracy (see subsection~\ref{subsec:truncation}). The mitigation of image truncation translates into smaller low-order aberrations. Profile variation coupled with spot truncation is very detrimental to LGS low-order wavefront sensing. We can say that a sodium density monitor is useful to mitigate the wavefront sensing risks (since it reduces the effect of truncation) but not sufficient to avoid a reference WFS, based on NGS, to measure the LGS spurious aberrations. The total RMS WFE related to sodium monitor data are also wavefronts with the minimum defocus term. In principle, we could avoid a large spot truncation and mitigate the wavefront sensing risks by using the focus error measured by the reference WFS. This is used as parameter to control the LGS focus mechanism in order to change the LGS reference altitude.
\section{Conclusions}
This paper reported some laboratory experiments realised to verify the performance of LGS wavefront sensing with a SH-WFS under different working conditions. The precision of the WFS using WCoG centroiding was evaluated across a range of SNRs representative of LGS AO facilities with current laser technology for 40-m class telescopes. Three LGS features mainly affect wavefront
sensing:
\begin{itemize}
\item The LGS image truncation introduces spurious aberrations whose magnitude depends on the level of spot truncation. These aberrations mainly fall within the first 60 Zernike modes and can be controlled by a reference WFS based on NGSs. To mitigate the wavefront sensing risks, the focus error, mesured by the reference WFS, can be used to control the LGS reference altitude avoinding a large image truncation. 
\item The sodium layer temporal evolution has the same impact of image truncation on wavefront sensing accuracy. It can be controlled by the reference WFS on timescale longer than atmospheric turbulence but short enough to monitor the sodium density profile variation. 
\item The LGS image under-sampling is critical. It could be useful to avoid spot truncation in a SH-WFS but a mitigation error strategy has to be implemented. 
\end{itemize}
In brief, the important achievements of the LGS WFS experiment are:
\begin{itemize}
\item	Supporting the development of a numerical simulation code for modelling the MAORY instrument for ELT. The numerical code is a much more flexible tool for designing the instrument, however the support given by this experimental work has been essential.
\item	Supporting the definition of the architecture of the MAORY instrument, concerning the requirements for the NGS wavefront sensing sub-system that supplement the LGS WFS.
\item	Supporting the definition of mitigation strategies for spot truncation / under-sampling which affect the performance of LGS WFS on ELTs.
\end{itemize}

\section*{Acknowledgements}
This paper used sodium layer data ~\citep{pfrommer2014high} kindly provided by Paul Hickson (University of British Columbia, Canada). The authors would like to thank the Italian Ministero dell'Istruzione, dell'Universit\`a e della Ricerca and the European Community Framework Program 7 (OPTICON project, grant agreement no. 312430) which supported this research project.


\bibliographystyle{mnras}
\bibliography{Pap} 


\bsp	
\label{lastpage}
\end{document}